\documentstyle[multicol, prb,aps,epsfig]{revtex}

\begin{document}
\date{\today}
\title{Crystallization of hard spheres under gravity}
\author{\bf Yan Levin}
\address{\it Instituto de F\'{\i}sica, Universidade Federal
do Rio Grande do Sul\\ Caixa Postal 15051, CEP 91501-970, 
Porto Alegre, RS, Brazil\\ 
{\small levin@if.ufrgs.br}}
\maketitle
\begin{abstract}

We present a simple argument to account for crystallization
of hard spheres under the action of a gravitational field.
The paper attempts to  bridge the gap  between two communities
of scientists, one working on granular materials and the 
other on inhomogeneous liquid state theory.

\end{abstract}

\centerline{{\bf PACS numbers:} 64.60.-i; 68.45.-v; 05.20-y;}
\bigskip


\section{Introduction}

Sedimentation of colloidal particles in a gravitational field
has been the subject of continued interest since the early
work of Jean Perrin. The external 
gravitational field breaks the translational symmetry,
producing an altitude dependent density profile. For a 
dilute suspension in which the interactions between the colloidal
particles can be neglected, the density profile takes
a simple exponentially decreasing form.  In fact it was this
observation which has first allowed Perrin in 1910 
to obtain an estimate of the Avogadro number and the 
Boltzmann constant\cite{per10}.  This early
work has stimulated much of the modern development of inhomogeneous
fluid theory.  

A few years ago it has been observed in Monte Carlo simulations
that a system of hard spheres can become unstable under the action
of a sufficiently strong gravitational field\cite{bib94}.  This instability
appeared as a sudden crystallization of the bottom layers of the
sample confined to a tree dimensional rectangular box.  In order
to understand this transition the authors appealed to the Weighted Density
Functional theory (WDF).  Unfortunately the WDF is quite computationally
demanding and is not very physically transparent. Perhaps
it is for this reason that this work went unnoticed by the community
of physicists studying granular materials\cite{hon99}.  Thus, the fluid-solid
condensation under the gravitational field 
has been recently rediscovered in the context of
granular materials.  Here the instability appeared 
as the break down of the Enskog kinetic theory for elastic hard
spheres\cite{ens22}, which Hong associated with the formation of a solid
layer at the bottom of a sample \cite{hon99}.  We must stress, however,
that the identification of the break down point of the Enskog equation
with crystallization is far from obvious.  In particular,
it is possible to show that at equilibrium the Enskog theory
is equivalent to the Local Density Approximation (LDA) \cite{hon99}.
Within the liquid state theory it is well known that the LDA 
cannot be used to obtain reliable information 
about the fluid-solid coexistence. The reason for this is that the 
LDA fails to  account for strong density variations
in a highly structured phase, such as the  
solid \cite{nor80,tar84}. Furthermore, in the absence of
an external field, the LDA for
hard spheres reduces to the usual
Carnahang-Starling equation of state \cite{car69}, which carries no 
information about the entropy driven crystallization of
a homogeneous hard sphere fluid. It is difficult,
therefore,
to understand how a theory which knows absolutely nothing about the
existence of an instability in a homogeneous system, 
suddenly develops this ``knowledge'' in a 
presence of a gravitational field. Thus, it is  dubious that 
the break down of the LDA  --- and, therefore, 
of the Enskog theory ---can be associated with a phase
transition, and not with the failure of the 
approximations underlying the theory.

In view of all this, it seems  worth while to present a simple
argument to account for  crystallization of hard spheres under the
action of an external gravitational field.  

\section{Lattice gas in a gravitational field}

Before entering into the discussion of a phase transition in the
system of hard spheres, we shall explore a much simpler model
of fluid and granular matter \cite{sel00} in a gravitational
field --- a non-interacting lattice gas.  
Consider a three dimensional rectangular volume $V=L^2H$,
divided into cubic cells of volume $a^3$.  Each cell can be occupied
by at most one ``cubic'' particle of mass $m$ and volume $a^3$ equal to
that of a unit cell.  The constant gravitational field of strength $g$
acts in the $-z$ direction. Since the particles do not
interact, in the thermodynamic limit,  
it is a simple matter to write an exact Helmholtz
free energy functional, 
\begin{equation}
\label{1}
\frac{\beta a^2 F[\sigma(z)]}{L^2}=\sum_{z=0}^\infty \left\{
\left[1-\sigma(z)\right]\ln\left[1-\sigma(z)\right]+\sigma(z)\ln \sigma(z)
+\beta m a g z\,\sigma(z) \right\}\;,
\end{equation}
where $\beta=1/k_B T$, and 
$\sigma(z)$ is the average occupation of the row $z$. 
The equilibrium  density profile can be obtained by minimizing 
the functional (\ref{1}), 
subject to the constraint of conservation of the total number
of particles $N$,
\begin{equation}
\label{2}
\sum_{z=0}^{\infty} \sigma(z)=Na^2/L^2 \equiv n \;.
\end{equation}
Performing a simple
calculation we find,
\begin{equation}
\label{3}
\sigma(z)=\frac{1}{1+\mbox{e}^{\beta m a g z+ \beta \mu }}\;,
\end{equation}
where the chemical potential $\mu$
is the Lagrange multiplier associated with the
constraint (\ref{2}).  It is convenient to define the reduced
parameters, $\beta^*=\beta m g a$ and $\mu^*=\mu/m g a$. In
the limit $-\beta \mu \gg 1$ the reduced chemical potential is
very well approximated by $\mu^*=-n$.  The density profile
now takes a particularly simple form,
\begin{equation}
\label{4}
\rho(z)=\frac{m}{a^3}\frac{1}{1+\mbox{e}^{\beta^* (z- n) }}\;.
\end{equation}
On the other hand the static pressure inside the lattice gas
satisfies,
\begin{equation}
\label{5}
\frac{d P(z)}{dz}=-g a \rho(z)\;,
\end{equation}
where to simplify the calculations we have passed to the continuum limit.
Eq.~\ref{5} can now be integrated yielding,
\begin{equation}
\label{6}
P(z)=\frac{m g}{a^2}\left(n+\frac{\ln\left[1+\mbox{e}^
{\beta^*( z-n)}\right]}{\beta^*}-z \right)\;.
\end{equation}
We note that the pressure is a uniformly decreasing function of the
height $z$.  At high altitudes --- where the average density is small and
the excluded volume constraint is irrelevant --- the density profile
reduces to the exponentially decreasing form first obtained by Perrin.

\section{The hard sphere fluid in the gravitational field}

It is evident from the analytic expressions derived above that the
lattice gas --- in spite of its hard core repulsion  expressed
through the constraint of one particle per lattice site --- is completely
stable for  all temperatures. It is reasonable to suppose that
to zeroth order the density profile of a hard sphere
fluid in a gravitational field should also be
well approximated by 
the Eq.~(\ref{4}). In fact it
has been found that the Fermi-like distribution (\ref{4}), provides
an excellent fit for various granular materials including the 
hard spheres \cite{hay97,cle91}.  

Unlike the non-interacting lattice gas, the homogeneous
system of hard spheres 
is known to undergo a structural fluid-solid transition \cite{cha88} when the
bulk pressure reaches $P_b \approx 14 k_BT/a^3$.  
This entropic transition was first observed in Monte Carlo simulations
almost  half a century ago \cite{ald57}, and has since been confirmed by
various theoretical methods \cite{ram79,tar84,bau85}. In the presence of a gravitational
field the pressure inside the  suspension varies uniformly
with maximum value located at $z=0$.  In the limit $\beta^* n \gg 1$,
the pressure at the bottom is well approximated by $P(0)=m g n/a^2$.  
Therefore,  the first layer will crystallize when,
\begin{equation}
\label{7}
P(0)=\frac{m g n}{a^2} \approx \frac{14 k_BT}{a^3}\;,
\end{equation}
or when 
\begin{equation}
\label{8}
T<T_c=\frac{m g a n }{14 k_B}\;.
\end{equation}
This is a discontinuous first order transition.  If the temperature
is lowered further, additional fluid layers will solidify and
a growing crystal will coexist with a diminishing fluid phase.  The number
of layers in the crystal can be determined by comparing $P(z)$ with 
$P_b$. We note that in the region of phase transition,
\begin{equation}
\label{9}
\beta^* n \approx -\beta^* \mu^* \approx 14 \gg 1\;,
\end{equation}
so that all the approximations adopted above are indeed justified.

We have presented a simple argument for crystallization of hard spheres
in a gravitational field.  This transition is nothing more than
a realization of the well known entropically driven liquid-solid phase
separation of a homogeneous hard sphere fluid. The presence of a
gravitational field leads to an altitude dependent pressure, which 
increases with the depth of suspension.  
If the temperature is reduced bellow the critical value $T_c$,
the pressure on the bottom layers becomes so large as to lead to
their crystallization.

A comment should be made about the real granular materials.  
For macroscopic particles, 
temperature is an irrelevant parameter.  Thus, in order
to study dynamic equilibrium or steady-state,  
kinetic energy must be supplied to 
the system through, say, a vibrating bed\cite{cle91}. 
If the collisions between the particles
are elastic, in the disordered (fluid) state we can 
think of the  kinetic energy as a form of  
``temperature''.  Appealing to the same argument as
above, we  reach the conclusion that the
granular materials also undergo phase separation
under the action of a gravitational field.
Some care, however, must  be taken with this argument.  
It is well known that unlike for fluids, the pressure
inside a container filled with packed granular  matter 
saturates at some value \cite{wol97}. The excess pressure
is transfered to the walls and is equilibrated by friction forces.
At high packing fractions, where the solidification takes place,
the stress is not uniform  but is
concentrated along force lines. This unusual characteristics
can significantly affect the location of the phase transition
in real granular materials.

\section{Acknowledgements}   
I would like to acknowledge many interesting conversations with
J. Arenzon. This work was supported in part by  
Conselho Nacional de
Desenvolvimento Cient{\'\i}fico e Tecnol{\'o}gico (CNPq),  Financiadora 
de Estudos e Projetos (FINEP).


\end{document}